\documentstyle[12pt,epsfig]{article}
\title{Finite-dimensional analogs of string $s \leftrightarrow t$
duality and pentagon equation}
\author{Igor G. Korepanov\thanks{Permanent address:
South Ural State University, 76 Lenin ave., Chelyabinsk 454080, Russia.
E-mail address: igor@prima.tu-chel.ac.ru} \quad
and \quad Satoru Saito\thanks{E-mail address: saito@phys.metro-u.ac.jp}\\
\small $\matrix{\noalign{\vskip3mm}
\hbox{Department of Physics, Tokyo Metropolitan University}\cr
\hbox{Minami-Ohsawa, Hachioji, Tokyo 192-0397, Japan}}$}
\date{December 1998}

\def\be{\begin{equation}}
\def\ee{\end{equation}}
\def\d{{\,\rm d}}

\begin{document}
\maketitle

\baselineskip=1.5\baselineskip

\begin{abstract}
We put forward one of the forms of functional pentagon equation
(FPE), known from the theory of integrable models,
as an algebraic explanation to the phenomenon known in physics
as $s\leftrightarrow t$ duality. We present two simple geometrical
examples of FPE solutions, one of them yielding in a particular case
the well-known Veneziano expression for 4-particle amplitude.
Finally, we interpret our solutions of FPE in terms of relations
in Lie groups.
\end{abstract}

\section{Introduction}

\begin{figure}
\begin{center}
\psfig{file=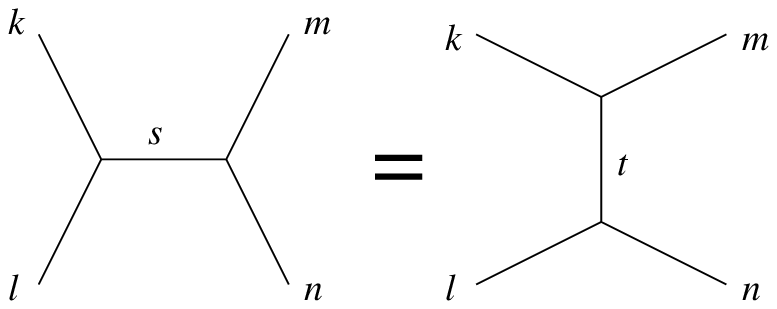}
\end{center}
\vspace*{5mm}
\caption{$s \leftrightarrow t$ duality}
\label{fig s-t duality}
\end{figure}

One of the most characteristic properties of scattering in string theory,
as well as in some strong interaction processes, is the so-called
$s \leftrightarrow t$ duality. Schematically, it can be represented
as in Figure~\ref{fig s-t duality}.

We assume here the point of view that the 4-point amplitudes
$A(k,l,m,n)$, where $k,l,m,n$ are some `quantum numbers', in both
sides of Figure~\ref{fig s-t duality} are obtained out of two 3-point
amplitudes $A(k,l,m)$ by integrating away the quantum number corresponding
to the internal line, so that analytically the duality looks like
\be
\int A(k,l,s)A(s,m,n) \d\mu(s)=\int A(k,m,t)A(t,l,n) \d\mu(t),
\label{eq duality}
\ee
where $\mu$ is some measure. In~(\ref{eq duality}) the 3-point amplitude
is assumed to be symmetric in all its arguments. If it is not, the
formula~(\ref{eq duality}) must be written more carefully, as we will
see below in section~\ref{sec k's}.

As is known, the $s \leftrightarrow t$ duality results in a dramatic
reduction of the number of Feynman diagrams: any two diagrams with the same
numbers of external lines and cycles are equivalent.

\begin{figure}
\begin{center}
\psfig{file=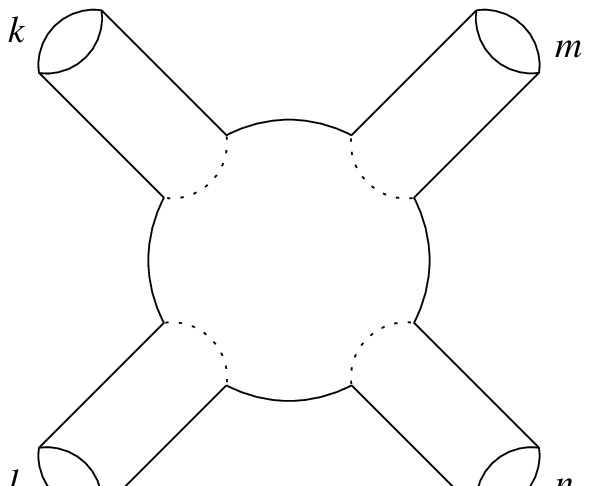}
\vspace*{8mm}
\end{center}
\caption{`Topological' explanation of $s\leftrightarrow t$ duality: this
diagram (string world sheet) corresponds to both l.h.s.\ and r.h.s.\ of
Figure~\protect\ref{fig s-t duality}}
\label{fig s-t string}
\end{figure}

What mathematical structures form the basis of such duality? Of course,
within the usual string theory it can be explained `geometrically' by
saying that to the two sides of Figure~\ref{fig s-t duality} corresponds in
fact the same `string diagram' (Figure~\ref{fig s-t string}).
Suppose, however, that we want to have a general algebraic mechanism for
$s \leftrightarrow t$ duality irrespective of such pictures and
hopefully providing new possibilities for constructing string-like
theories.

String theory has intimate connections with many fields of mathematics.
We are mostly interested in its relations with integrable models.
So, let us mention here the string--soliton
correspondence~\cite{saito1,saito2,sogo} and
the fact that the string amplitudes satisfy the Yang--Baxter
equation~\cite{saito3}.

There exist, however, different kinds of fundamental equations responsible
for integrability. The most important is believed to be the tetrahedron
equation (TE), which deals with $2+1$-dimensional integrability: the
quantum TE~\cite{zamolodchikov} for quantum models and the functional
TE~\cite{kashaev1,KKS} for both classical and quantum models. At the same
time, the different equations are strongly
connected with one another. In particular, the tetrahedron equation
is connected with the {\em pentagon equation}~\cite{KS}.

In this paper we argue that the mathematical structure responsible for
$s \leftrightarrow t$ duality is the functional pentagon equation~(FPE).
We demonstrate it on simple examples. As is known, strings have close
relations with infinite-dimensional groups. Nevertheless, we believe
it is natural to start from FPE solutions related to finite-dimensional
Lie groups. So, the modest aim of this paper is to demonstrate that there
can exist some algebraic mechanism for $s \leftrightarrow t$ duality based
on FPE solutions, and if we can pass from finite-dimensional to
infinite-dimensional groups (which looks very plausible), we will obtain
wide range of new string-like theories.

Below, in section~\ref{sec pentagon} we explain what is the FPE and how
it arises naturally when dealing with $s \leftrightarrow t$ duality.
In sections~\ref{sec lengths} and~\ref{sec k's} we present two simple
geometric constructions for 3-point amplitudes that obey duality.
The amplitude of section~\ref{sec k's} generalizes
the well-known Veneziano 4-particle amplitude.
Finally, in section~\ref{sec discussion} we show that our
constructions can be described algebraically in terms of relations in Lie
groups, namely the group of movements of euclidean plane 
and the Heisenberg group.

\section{Functional pentagon equation}
\label{sec pentagon}

Our idea of a duality mechanism is very simple. Suppose that, for any
fixed quantum numbers on the external lines of both sides in
Figure~\ref{fig s-t duality},
there exists some correspondence law $f\colon\; t\mapsto s$ such that
\be
t=f(s)\; \Rightarrow\; A(k,l,s)A(s,m,n)\d\mu(s)
 = A(k,m,t)A(t,l,n)\d\mu(t).
\label{eq s-t transformation}
\ee
Here it is implied tacitly that function $f$ also depends
on the `outer' variables $k,l,m$ and~$n$.

\begin{figure}
\begin{center}
\psfig{file=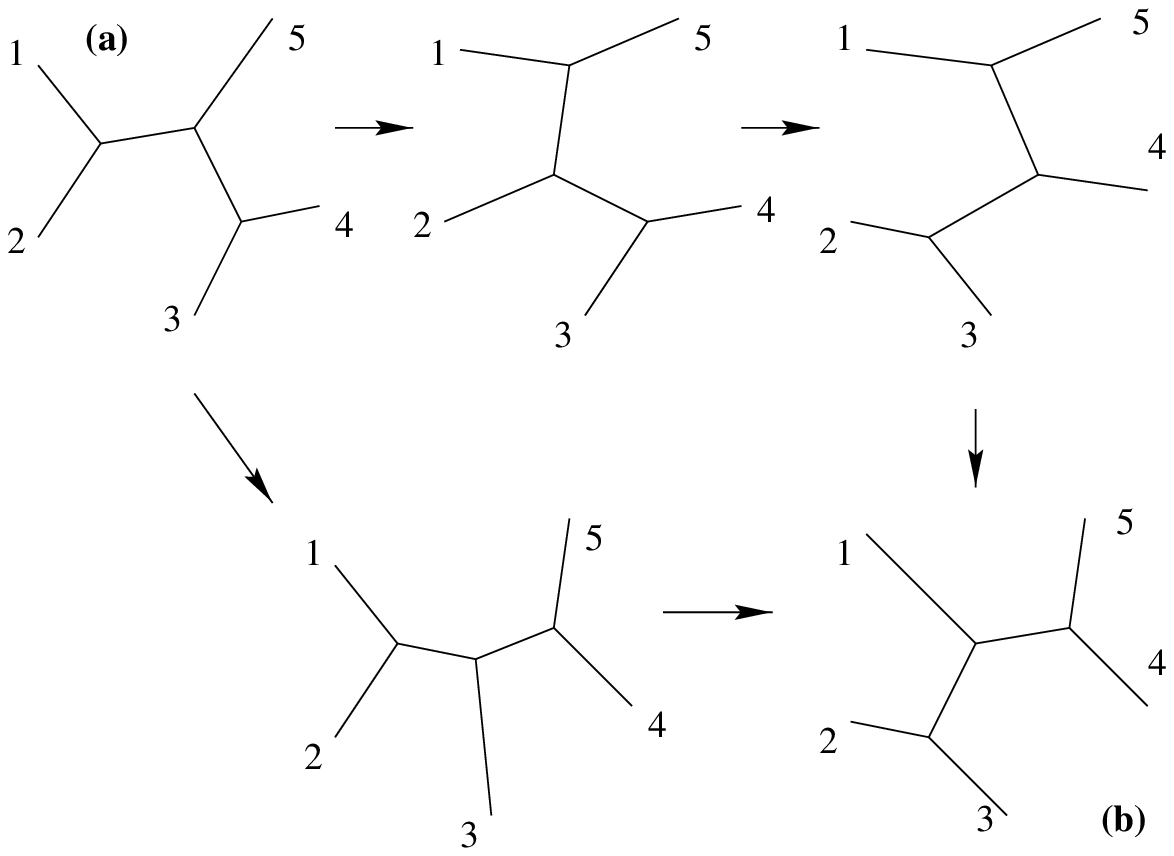}
\vspace*{5mm}
\end{center}
\caption{Two ways of transforming quantum numbers}
\label{fig s-t pentagon}
\end{figure}

It is clear that (\ref{eq s-t transformation}) is sufficient for
(\ref{eq duality}) to hold. Consider now the scattering diagram~(a)
in Figure~\ref{fig s-t pentagon}
and transform it into the diagram~(b)
in two ways, as shown in the Figure. Suppose that some quantum numbers
have been attached to {\em all\/} lines of diagram~(a) (including the internal
ones). Using the function~$f$, we get quantum numbers for diagram~(b)
as well. It is very natural to require that
the function~$f$ satisfy the compatibility condition: {\em two sequences of
transformations in Figure~\ref{fig s-t pentagon} must result in the
same quantum numbers for diagram~(b)}. And this compatibility condition
is nothing but some version of functional pentagon equation.

\begin{figure}
\begin{center}
\psfig{file=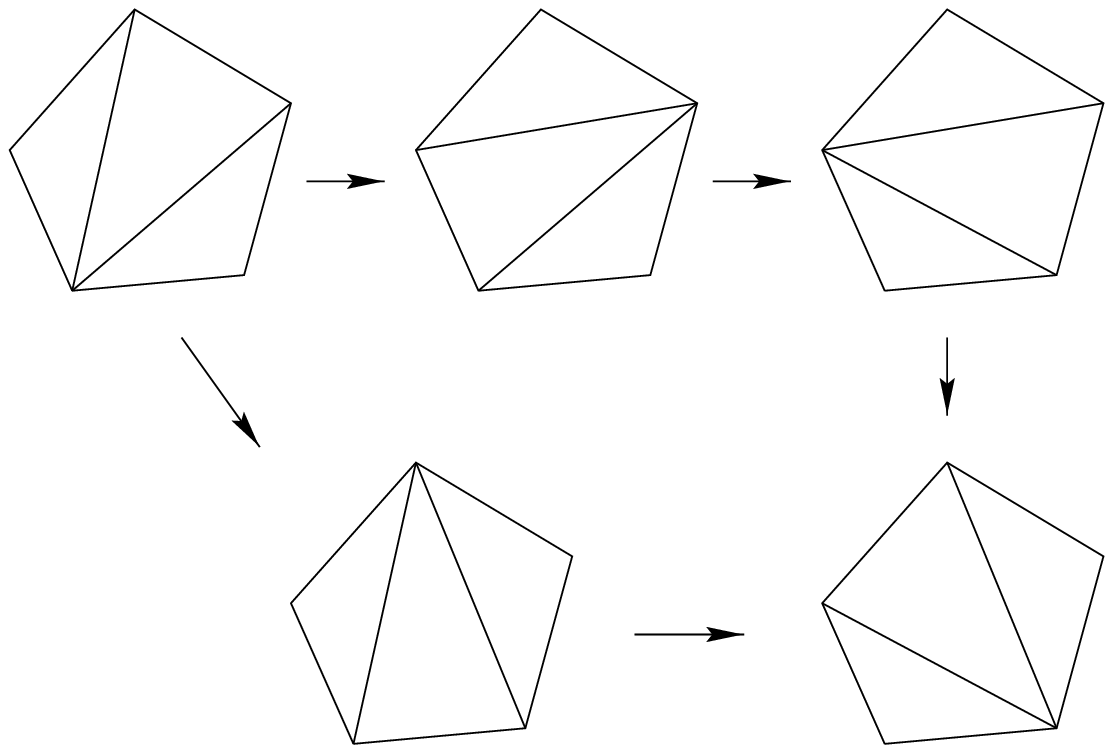}
\vspace*{5mm}
\end{center}
\caption{The pentagon equation}
\label{fig pentagon}
\end{figure}

To see where the pentagon is, let us draw a `Poincar\'e dual' for
Figure~\ref{fig s-t pentagon}
as Figure~\ref{fig pentagon}.
Here the vertices of Figure~\ref{fig s-t pentagon} are represented as
triangles, and to a transform of the type of Figure~\ref{fig s-t duality}
corresponds deleting of a diagonal of a quadrilateral and replacing it
with the other diagonal.

Note that the variables (quantum numbers) are attached to the {\em edges\/}
(sides and diagonals) of the pentagon. There exist also other versions
of FPE where variables belong e.g.\ to the triangles themselves,
see~\cite{KS,kashaev2}.

\section{Geometric duality for edge lengths}
\label{sec lengths}

\begin{figure}
\begin{center}
\psfig{file=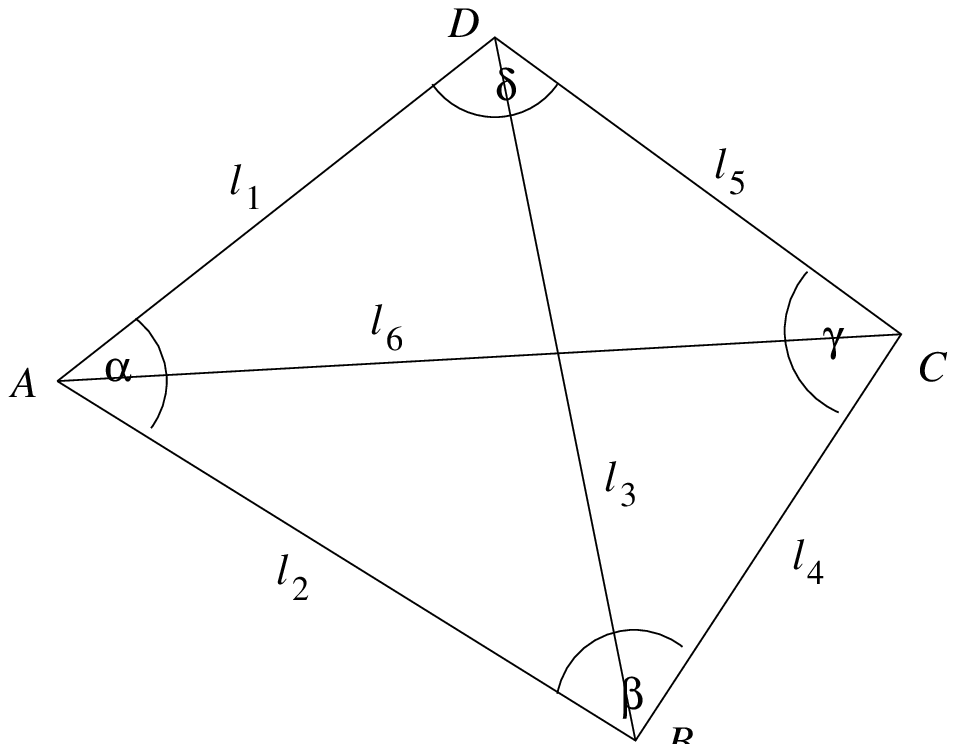}
\vspace*{5mm}
\end{center}
\caption{Edge lengths as quantum numbers}
\label{fig lengths}
\end{figure}

The geometrical picture of Figure~\ref{fig pentagon} suggests at once
a possibility for choosing function~$f$. Namely, let us draw, as
Figure~\ref{fig lengths}, the
`Poincar\'e dual' of Figure~\ref{fig s-t duality} on the euclidean plane
and take the {\em lengths of edges\/} as `quantum numbers'. If the
lengths $l_1,l_2,l_3,l_4$ and $l_5$ of edges in Figure~\ref{fig lengths}
are given, then $l_6$ is determined from the equation
\be
S_{ABD}+S_{BCD}=S_{ABC}+S_{ACD},
\label{eq areas}
\ee
where $S_{\ldots}$ is the area of the corresponding triangle expressed
through the lengths of its sides. For example,
$$
S_{ABD}=S(l_1,l_2,l_3)={1\over 4}\,\root\of{(l_1+l_2+l_3)(l_2+l_3-l_1)
(l_3+l_1-l_2)(l_1+l_2-l_3)}.
$$

The fact that such a transformation $f\colon\; l_3\mapsto l_6$ obeys the
pentagon equation is evident from geometrical argument (a pentagon in which
the lengths of all sides and two diagonals are given is a `rigid body'
where distance between any two points is fixed and does not depend on a
chain of algebraic transformations we have used to calculate it).

Consider the obvious relation
\be
\d (\alpha+\gamma)=-\d (\beta+\delta)
\label{eq angles}
\ee
for the angles in Figure~\ref{fig lengths}, and let the sides of the
quadrilateral~$ABCD$ be fixed and only its diagonals vary. Using formulae
like
$$
\sin\alpha={2S_{ABD}\over l_1l_2}, \quad \cos\alpha=
{l_1^2+l_2^2-l_3^2\over 2l_1l_2},
$$
we will find
$$
\d\alpha=-{\d\cos\alpha\over \sin\alpha}={l_3 \d l_3\over 2S_{ABD}}
$$
and similarly for $\beta,\gamma$ and $\delta$. Substituting these angle
differentials into (\ref{eq angles}) and taking (\ref{eq areas}) into
account, it is not hard to derive the relation
\be
{l_3\d l_3\over S_{ABD}\cdot S_{BCD}}=
-{l_6\d l_6\over S_{ABC}\cdot S_{ACD}}.
\label{eq ldl/SS}
\ee

The relation (\ref{eq ldl/SS}) together with (\ref{eq areas}) suggests
the following form for 3-point amplitude $A(l_1,l_2,l_3)$ and measure
$\d\mu(l)$ satisfying the condition~(\ref{eq s-t transformation}):
\be
A(l_1,l_2,l_3)={e^{\lambda S(l_1,l_2,l_3)}\over S(l_1,l_2,l_3)},
\label{eq lengths amplitude}
\ee
where $\lambda$ is an overall arbitrary constant, and
\be
\d\mu(l)=l\d l.
\label{eq lengths measure}
\ee

As for the minus sign in (\ref{eq ldl/SS}), its role becomes clear when
we choose the integration path in $l_3$ and/or $l_6$. Here some freedom
seems to exist. If we regard the lengths as complex variables, the
equation (\ref{eq areas}) determines, for fixed $l_1,l_2,l_4$ and $l_5$,
some Riemann surface whose points are pairs $(l_3,l_6)$. So, probably,
some cycles on that surface can be taken as integration contours. Here
we will not go that far, but just na\"\i vely assume $ABCD$ to be a convex
quadrilateral in a usual real euclidean plane,
and let $l_3$ change from its minimal value compatible
with this assumption (and with given $l_1,l_2,l_4$ and $l_5$)
to its maximal value. Then it is clear that $l_6$ changes from {\em its
maximal value to its minimal value}. If we reverse the direction of
integration in $l_6$, the minus sign disappears, and finally the
formula~(\ref{eq duality}) acquires the form
$$
\intop_{(l_3)_{\rm min}}^{(l_3)_{\rm max}}
{e^{\lambda S(l_1,l_2,l_3)}\over S(l_1,l_2,l_3)}
{e^{\lambda S(l_3,l_4,l_5)}\over S(l_3,l_4,l_5)}
\,l_3 \d l_3 =
\intop_{(l_6)_{\rm min}}^{(l_6)_{\rm max}}
{e^{\lambda S(l_1,l_5,l_6)}\over S(l_1,l_5,l_6)}
{e^{\lambda S(l_2,l_4,l_6)}\over S(l_2,l_4,l_6)}
\,l_6 \d l_6.
$$

\section{Geometric duality for angular coefficients---a generalization of
Veneziano amplitude}
\label{sec k's}

\begin{figure}
\begin{center}
\psfig{file=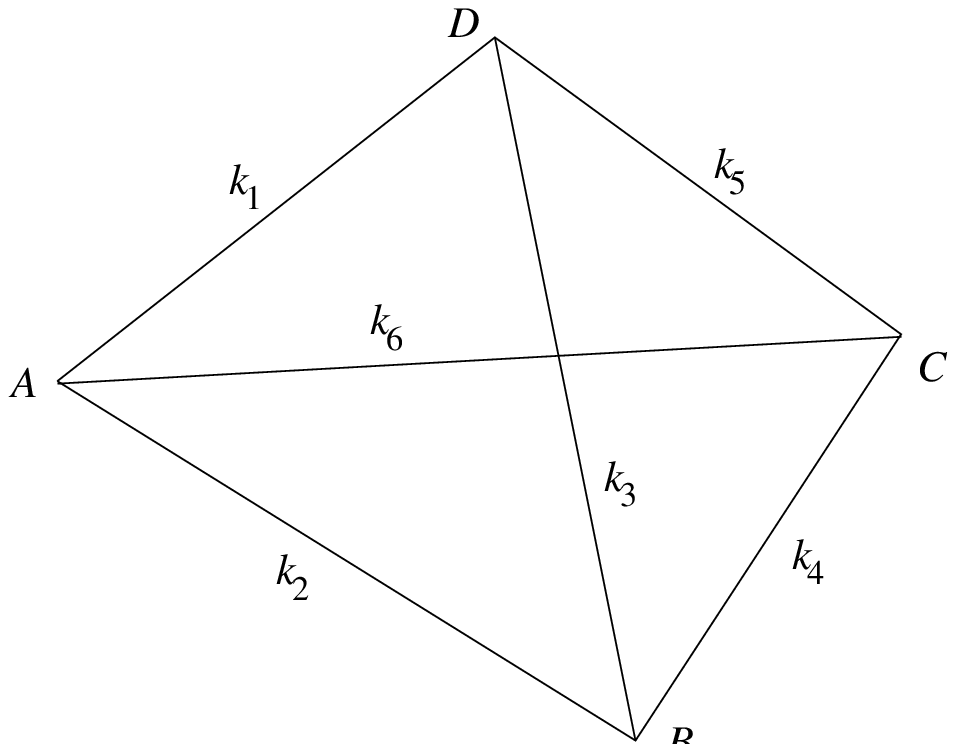}
\vspace*{5mm}
\end{center}
\caption{Angular coefficients as quantum numbers}
\label{fig k's}
\end{figure}

Figure~\ref{fig pentagon} suggests in fact one more choice of a
transformation satisfying the FPE. Let us take as a quantum number
the {\em angular coefficient}~$k$ of a given edge (if the ends of the edge
are $(x_1,y_1)$ and $(x_2,y_2)$ in some fixed frame of reference, not
necessarily orthogonal, then $k=(y_2-y_1)/(x_2-x_1)$). It is not hard to see
that if values $(k_1,k_2,k_3,k_4)$ and $k_5$ in Figure~\ref{fig k's}
are given, then $k_6$ is determined uniquely.
Let us write the formula for finding~$k_6$ in the following form:
\be
{k_3-k_2\over k_3-k_1}\cdot {k_3-k_5\over k_3-k_4}=
{k_6-k_2\over k_6-k_4}\cdot {k_6-k_5\over k_6-k_1}.
\label{eq k's}
\ee
The structure of this relation is
\be
F(ABD)\cdot G(BCD)=H(ABC)\cdot K(ACD),
\label{eq *}
\ee
by which we mean that
the l.h.s.\ is the product of two expressions corresponding to triangles
$ABD$ and $BCD$ respectively, while the r.h.s.\ corresponds in a similar way
to triangles $ABC$ and~$ACD$. So, relation~(\ref{eq k's}) is
similar to~(\ref{eq areas}), although it is `multiplicative' rather than
`additive'.

Note that (a)\,equation~(\ref{eq k's}) has also the solution $k_6\equiv k_3$
which we are not interested in. Let us agree that we have
rejected that solution; (b)\,the transformation $f\colon\; k_3\mapsto k_6$,
given $k_1,k_2,k_4$ and $k_5$, is an involution.

The fact that the transformation $f$ obeys the pentagon equation follows
from the fact that a pentagon for which the angular coefficients of all
sides and two diagonals are given is determined uniquely up to
a similarity and a shift.

Remarkably, there exist two more multiplicative relations yielding the
same dependence $f\colon\; k_3\mapsto k_6$ and having the same
structure~(\ref{eq *}):
\be
{k_1-k_2\over k_1-k_3}\cdot {k_4-k_3\over k_4-k_5}=
{k_4-k_2\over k_4-k_6}\cdot {k_1-k_6\over k_1-k_5}
\label{eq *2}
\ee
and
\be
{k_2-k_1\over k_2-k_3}\cdot {k_5-k_3\over k_5-k_4}=
{k_2-k_6\over k_2-k_4}\cdot {k_5-k_1\over k_5-k_6}.
\label{eq *3}
\ee

For the analog of formula~(\ref{eq ldl/SS}) we can take
\be
{\d k_3\over (k_3-k_2)(k_3-k_5)}=-{\d k_6\over (k_6-k_2)(k_6-k_5)}.
\label{eq dk}
\ee
The reader can verify that (\ref{eq dk}) follows from (\ref{eq k's}),
provided $k_6\not\equiv k_3$.

An important feature of formulae (\ref{eq k's}--\ref{eq dk}) is that
they remain valid after a M\"obius (rational) transformation of all $k$'s:
$$
k_j\mapsto {ak_j+b\over ck_j+d},\quad j=1,\ldots,6.
$$
Such transformations correspond just to another choice of coordinate axes
for Figure~\ref{fig k's}.

The most general formula of type~(\ref{eq duality}) that we can obtain from
(\ref{eq k's}--\ref{eq dk}) results from raising all terms in
(\ref{eq k's}) to some degree~$\alpha$, in (\ref{eq *2}) to some
degree~$\beta$, in (\ref{eq *3}) to some degree~$\gamma$ and multiplying
all together with corresponding
terms of~(\ref{eq dk}). Using the fact that
$$
k_3=k_2\; \Leftrightarrow \; k_6=k_5\quad \hbox{and} \quad
k_3=k_5\; \Leftrightarrow \; k_6=k_2,
$$
we can choose e.g.\ a curve joining $k_2$ and $k_5$ as the integration path,
and write the final formula as
\begin{eqnarray}
\intop_{k_2}^{k_5} A(k_3,k_1,k_2 | -\alpha,\beta,-\gamma)\,
A(k_3,k_4,k_5 | -\alpha,-\beta,\gamma) \d k_3 \nonumber\\
= \intop_{k_2}^{k_5} A(k_6,k_4,k_2 | -\alpha,\beta,\gamma)\,
A(k_6,k_1,k_5 | -\alpha,-\beta,-\gamma) \d k_6,
\label{eq duality k's}
\end{eqnarray}
where
\be
A(l,m,n | \lambda,\mu,\nu)= (l-m)^{\lambda-\mu} (m-n)^{\mu-\nu}
(n-l)^{\nu-\lambda-1}.
\label{eq amplitude k's}
\ee

Both sides of (\ref{eq duality k's}) generalize the
well-known Veneziano expression~\cite{veneziano,GSW} for the
4-point amplitude. To see this, let us put
$$
k_1=k_4=\infty,\quad k_2=0,\quad k_5=1.
$$
Then, for example, the l.h.s. of (\ref{eq duality k's}) yields,
up to a constant multiplier,
$$
\int_0^1 k_3^{\alpha-\gamma-1} (k_3-1)^{\alpha+\gamma-1} \d k_3,
$$
which coincides with the Veneziano amplitude up to the obvious
change of notations.

It will be certainly of big interest to compare the {\em five}-point
and, more generally, $N$-point
amplitudes that can be obtained in such way with those in classical
papers~\cite{BR,virasoro,GS,KN2}.

\section{Discussion: a group-theoretical comment}
\label{sec discussion}

Let us explain why we believe that our constructions are related to Lie
groups. Consider two transformations acting on points $(x,y)$ of
a euclidean plane: shifting by~$a$ along the $x$~axis
\be
S(a)\colon\; (x,y)\mapsto (x+a,y)
\label{eq S(a)}
\ee
and rotation through the angle~$\phi$
$$
R(\phi)\colon\; (x,y)\mapsto (x\cos\phi-y\sin\phi, x\sin\phi+y\cos\phi).
$$
Then, the existence of a triangle with sides $l,m,n$ and {\em external\/}
angles $\alpha,\beta,\gamma$ means the equality
\be
R(\alpha)\circ S(n)\circ R(\beta)\circ S(l)\circ R(\gamma)\circ S(m)=
{\bf 1}.
\label{eq 6-term}
\ee

Note that the group of movements of a euclidean plane is three-parametric,
and that is why $\alpha,\beta$ and~$\gamma$ can be determined from given
$l,m,$ and~$n$. Similar to~(\ref{eq 6-term}) relations can be written also
for quadrilaterals and pentagons, and all the geometric constructions of
section~\ref{sec lengths} can be described in terms of such relations.

As for section~\ref{sec k's}, its constructions have nothing to do with
euclidean distance, so, in our opinion, here more relevant is the
three-parametric group generated by the transformations~$S(a)$~(\ref{eq S(a)})
and
$$
T(\kappa)\colon\; (x,y)\mapsto (x,y+\kappa x).
$$
Then the existence of a triangle whose sides have $x$-projections $a,b,c$
(where $a+b+c=0$) and angular coefficients $l,m,n$ is described by the
equality
\be
S(a)\circ T(l-m)\circ S(b)\circ T(m-n)\circ S(c)\circ T(n-l)={\bf 1}.
\label{eq 6-term'}
\ee

From the abstract point of view,
the group generated by $S(a)$ and $T(\kappa)$ is nothing but the Heisenberg
group. One can readily see this from the relation
$$
T(\kappa)\circ S(a)\circ T(-\kappa)\circ S(-a)=C(\kappa a),
$$
where
$$
C(a)\colon\; (x,y)\mapsto (x,y+a)
$$
is a central element for any~$a$.

We believe that studying relations of the
type~(\ref{eq 6-term},\ref{eq 6-term'}) in greater groups is the algebraic clue
for constructing wide generalizations of the string theory.

\subsubsection*{Acknowledgements}
One of the authors (I.K.) is glad to acknowledge the TMU Fellowship for
research
from Tokyo Metropolitan University, the hospitality of his host researcher
Professor S.~Saito, and the support from Russian Foundation for Basic
Research under grant no.~98-01-00895.

\end{document}